\documentclass[twocolumn]{revtex4}

\usepackage{amsmath,amssymb,bm,graphicx}

\begin{document}

\title{Assessing node risk and vulnerability in epidemics on networks}

\author{T. Rogers}

\affiliation{Centre for Networks and Collective Behaviour - University of Bath, Bath, BA2 7AY, UK}

\begin{abstract}
Which nodes are most vulnerable to an epidemic spreading through a network, and which carry the highest risk of causing a major outbreak if they are the source of the infection? Here we show how these questions can be answered to good approximation using the cavity method. Several curious properties of node vulnerability and risk are explored: some nodes are more vulnerable than others to weaker infections, yet less vulnerable to stronger ones; a node is always more likely to be caught in an outbreak than it is to start one, except when the disease has a deterministic lifetime; the rank order of node risk depends on the details of the distribution of infectious periods.
\end{abstract}
\maketitle

\section{Introduction}
Network structure has a profound influence over disease dynamics. There is growing acknowledgement of this fact in the epidemiology literature (see \cite{Keeling2005} for a review), where the question of how best to quantify and predict network effects is becoming a central challenge \cite{Pellis2014}. At the same time, network epidemic models are of considerable interest to the theoretical physics community \cite{Pastor2014}, serving as a canonical example of a non-equilibrium process, and providing fertile ground for the application of statistical mechanics techniques to new and interesting problems. One such technique is the cavity method: invented to tackle problems in statistical inference \cite{Pearl1982} and condensed matter physics \cite{Parisi1987}, this simple and versatile method has found application in areas as diverse as computer science \cite{Mezard2002} and random matrix theory \cite{Rogers2008,Rogers2009}. 

The cavity method was first applied to epidemiology in \cite{Karrer2010} (using the alternative moniker of `message-passing'), to calculate the expected time development of a network epidemic. The principle advance provided by the method in that work was the ability to model non-Markov epidemics, in which the time that a node remains infective (the lifetime distribution, or infectious period) is not a simple memoryless exponential distribution. Under a second alias of `belief-propagation', it has also been applied to the problem of tracing the most likely source of a disease outbreak \cite{Altarelli2014}. This work points to another advantage offered by the cavity method; that it is not restricted to calculating macroscopic quantities, but can provide information about the behaviour of individual nodes. Indeed, it is now well-recognised that population heterogeneity can play a very important role in disease outbreaks \cite{Lloyd2005}. 

In this article, we show how to calculate a measure of the infection risk that a particular node poses to the network, as well as the risk posed by the network to the node. These quantities are manifested in two different realisations of the cavity method, one with `upstream' cavities, the other `downstream'. The derivation of the cavity equations for the risk and vulnerability measures is given in the next section, followed by a discussion of their numerical solution. We then address the role of the lifetime distribution of the disease, which controls the probability of an outbreak occurring, but paradoxically has no effect on its final size. Finally, we show how the cavity method can be used to rank nodes according to their risk or vulnerability, and how these rankings exhibit curious dependencies on the disease dynamics. 

\section{Cavity method for risk and vulnerability}
Consider the spread of a disease over the nodes of a network. In a small period of time $dt$ each infected node has a chance $\beta \,dt$ of transmitting the infection to an uninfected neighbour, and a chance $\gamma(t)\, dt$ of recovering from the disease. Recovered nodes cannot be reinfected. This is an SIR epidemic with Poisson rate infections ($\beta$ is a positive constant) and general disease lifetime distribution $\gamma(t)$. 

Write $r_i$ for the risk that a major outbreak occurs if node $i$ is the sole initial infective. Thinking about the limit of very large networks, we make an artificial distinction between the local area around node $i$, and the rest of the network, which we call the `bulk'. We will say a major outbreak occurs if node $i$ succeeds in transmitting the infection a to non-zero fraction of the bulk. To compute an approximation to this quantity, we examine the spread of the disease away from the initial infection. 
\begin{figure}
\includegraphics{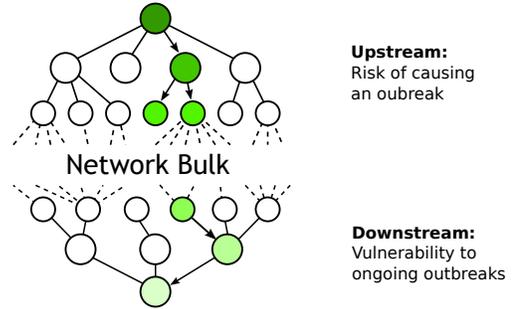}
\caption{Imagining the progress of the disease as a flow through the network, (\ref{cav_up}) considers the onward spread after removing an upstream neighbour (i.e. closer to the source of the infection), whilst (\ref{cav_down}) considers the chance of infection with a downstream neighbour removed. In both cases the tree approximation amounts to assuming that the up- and downstream neighbours are uniquely defined.}
\label{drawing}
\end{figure}

We begin by calculating the probability that $i$ infects all the vertices in some subset $J$ of its neighbours $\mathcal{N}(i)$, and no others. The fates of these nodes are correlated by their joint exposure to the random length of the infectious period of $i$; a long infectious period implies that nearly all neighbours will catch the disease, and the opposite holds if $i$ is infectious only for a very short time. If the infectious period has probability density $\gamma(t)$, then we may write 
\begin{equation}
\begin{split}
\mathbb{P}(i\to J)&=\int_0^\infty\gamma(t)\prod_{j\in J}(1-e^{-\beta t})\prod_{j\in\mathcal{N}(i)\setminus J}e^{-\beta t}\,dt\\
&=\int_0^\infty\gamma(t)(1-e^{-\beta t})^{|J|}e^{-\beta t(|\mathcal{N}(i)|-|J|)}\,dt\,.
\end{split}
\label{PiJ}
\end{equation}
After $i$ has infected some of its neighbours, the continued survival of the disease depends on those neighbours propagating the infection away from $i$. Summing over all possible collections of infected neighbours we find 
\begin{equation}
 r_i=1-\sum_{J\subset\mathcal{N}(i)}\mathbb{P}(i\to J)\,(1- r_J^{(i)})\,,
\label{rhoi1}
\end{equation}
where $ r_J^{(i)}$ is the probability of a major outbreak in a network where each $j\in J$ is initially infected, and \emph{node $i$ has been removed}. In words, this equation says that the disease starting at $i$ dies out if, whenever $i$ infects a set $J$, the disease starting at $J$ also dies out. 

The task now is to express $ r_J^{(i)}$ in terms of the $ r_j^{(i)}$. If the network contains cycles, then some nodes will be reachable from two or more members of $J$, yet only at most one can succeed in passing the infection. This gives rise to the inequality
\begin{equation}
1- r_J^{(i)}\geq \prod_{j\in J}(1- r_j^{(i)})\,.
\label{cav_approx}
\end{equation}
To compute a lower bound on $ r_i$ it will therefore suffice to assume equality in (\ref{cav_approx}). This approximation is equivalent to assuming that the network is large and tree-like. As we will see, the results remain surprisingly accurate for networks which do not fit this assumption very well. Indeed the ``unreasonable effectiveness'' of tree approximations appears to be a quite general phenomenon for network processes \cite{Melnik2011}.

Inserting (\ref{PiJ}) and (\ref{cav_approx}) into (\ref{rhoi1}) and assuming equality we can compute
\begin{equation}
\begin{split}
 r_i&=1-\int_0^\infty\gamma(t) \prod_{j\in\mathcal{N}(i)}\left(1-\big(1-e^{-\beta t}\big) r_j^{(i)}  \right)\,dt\\
&=1-\sum_{J\subset\mathcal{N}(i)}(-1)^{|J|}T_{|J|}\prod_{j\in J} r_j^{(i)}\,,
\end{split}
\label{r_eq}
\end{equation}
where
\begin{equation}
T_{n}=\int_0^\infty\gamma(t)(1-e^{-\beta t})^n\,dt\,.
\label{defT}
\end{equation}

The problem of computing the risk to the network posed by node $i$ has thus been transformed in to that of computing the risk posed by the neighbours of $i$ in the cavity graph. Repeating the same calculations for $ r_j^{(i)}$ allows us to derive a closed set of equations
\begin{equation}
 r_j^{(i)}=1-\sum_{L\subset\mathcal{N}(j)\setminus i}(-1)^{|L|}T_{|L|}\prod_{l\in L} r_l^{(j)}\,,\qquad\forall i,j.
\label{cav_up}
\end{equation}
These are the upstream cavity equations. The derivation requires the additional assumption that the risk posed by $l$ in the network with $i$ and $j$ both removed is well-approximated by $ r_l^{(j)}$, that is, we ignore any alternative path between $l$ and $i$. If we are able to solve these equations (more on this later), the results can be fed into equation (\ref{r_eq}) to provide us with an estimate of the probability of any given node causing a major outbreak. 

We may also ask how vulnerable node $i$ is to an outbreak that starts somewhere else in the network. Write $v_i$ for the probability that $i$ is eventually infected in the event of a major outbreak. To compute the cavity approximation for vulnerability, we trace the possible path of the disease downstream from the bulk to node $i$. The calculation is somewhat easier in this case, since each neighbour of $i$ has an independent chance to attempt to infect it. This means that the detail of the distribution of infectious periods no longer matters, only the overall chance of an infection, $T_1$ (known as the \emph{transmissibility}). The result, previously derived in \cite{Karrer2010}, is 
\begin{equation}
v_i=1-\prod_{j\in \mathcal{N}(i)}(1-T_1 v_j^{(i)})\,,
\label{vul}
\end{equation}
where the downstream cavity equations are 
\begin{equation}
v_j^{(i)}=1-\prod_{l\in \mathcal{N}(j)\setminus i}(1-T_1 v_l^{(j)})\,,\qquad\forall i,j.
\label{cav_down}
\end{equation}

The distinction between the up- and downstream cavity equations should hopefully be clear: see Figure \ref{drawing} for a cartoon illustration. It should also be noted that there is a close relationship between the cavity equations, and the equation for survival probability of a multi-type branching process \cite{Athreya1972}. This should not be surprising as the cavity equations are derived from considerations of the early development of the epidemic, where the branching process approximation is known to apply \cite{Ball1995}.

\section{Iteration and percolation}
\label{solution}

Typically, non-trivial solutions to the cavity equations in either direction cannot be found by hand. Fortunately, for general networks the cavity equations are stable under iteration and can be numerically solved to high accuracy with little computational cost. To do this efficiently, it is helpful to map to a new network encoding the relations between edges in the cavity equations. Each undirected edge $(i,j)$ spawns two nodes, $(i\to j)$ and $(j\to i)$. Between two nodes $e=(i\to j)$ and $e'=(i'\to j')$ in the new network, we draw a directed edge from $e$ to $e'$ if $r_{j'}^{(i')}$ is involved in the cavity equation for $r_{j}^{(i)}$, that is, if $i'=j$ and $j'\in\mathcal{N}(j)\setminus i$. 
\begin{figure}
\includegraphics{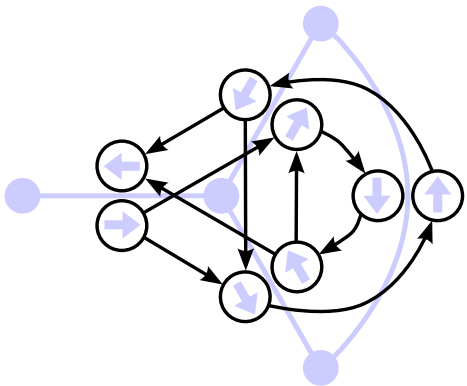}
\caption{Illustration of the non-backtracking network (dark, directed network) for a simple four-node network (pale, undirected network). Each edge in the underlying network spawns a pair of new nodes, one for each possible direction of infection, which are linked if they appear in the cavity equations.}
\label{cav_adj}
\end{figure}

The adjacency matrix $B$ of this network is known as the non-backtracking, or Hashimoto \cite{Hashimoto1989}, matrix. It has recently been shown to be of use in spectral clustering \cite{Krzakala2013}, and occurs in the calculation of the percolation threshold in sparse networks \cite{Karrer2014,Hamilton2014}. Using this construction, the upstream and downstream cavity equations can be rewritten as 
\begin{equation}
\bm{r}=U(\bm{r})\quad\text{and}\quad \bm{v}=D(\bm{v})\,,
\label{UDeq}
\end{equation}
where $U$ and $D$ are the vector functions 
\begin{equation}
\begin{split}
&U_e(\bm{x})=1-\sum_{E\subset\mathcal{N}(e)}(-1)^{|E|}T_{|E|}\prod_{e'\in E} x_{e'}\,,\\
&D_e(\bm{x})=1-\prod_{e' \in\mathcal{N}(e)}(1-T_1 x_{e'})\,.
\end{split}
\label{UD}
\end{equation}
Starting from the initial vector $x_e=1$ for all $e$, solutions to (\ref{UDeq}) can be found by repeatedly applying the maps $U$ or $D$. For $n\geq1$, the quantity $D^{n}_{(i\to j)}(\bm{1})$ describes the risk that the disease spreads at least distance $n$ from node $j$ in the absence of node $i$. Similarly, $U^n_{(i\to j)}(\bm{1})$ gives the risk that $j$ is infected if all nodes in the cavity network of distance greater than $n$ are themselves infected. Clearly these quantities are decreasing with $n$ and bounded from below by zero, hence the iteration scheme is guaranteed to converge. 

For efficient numerical implementation, the integrals (\ref{defT}) should of course be precomputed (indeed this can be accomplished analytically whenever the Laplace transform of $\gamma$ is known). It is possibly less obvious that often a substantial speed-up can be gained by also precomputing the explicit form of cavity equations themselves. That is, rather than having the computer loop through the subsets $E\subset\mathcal{N}(e)$ during each iteration step, an explicit iteration function can be procedurally generated from (\ref{UD}) before starting the main loop. 

Notice that the upstream and downstream equations (\ref{UDeq}) both admit zero as a solution, since it is always possible that the disease fails to spread. For some parameter values, the zero solution is unstable and the cavity equations admit a solution in $(0,1)$ corresponding to disease outbreak. The regimes of extinction and outbreak are separated by a percolation phase transition, and the critical parameter values can be determined by examining the stability of the maps $U$ and $D$ around zero. From (\ref{UD}) we compute 
\begin{equation}
\begin{split}
&\frac{\partial U_e}{\partial x_{e'}}=-\sum_{E\subset\mathcal{N}(e)}\mathbb{I}_{\{e'\in E\}}(-1)^{|E|}T_{|E|} \prod_{e''\in E\setminus e'} x_{e''}\,,\\
&\frac{\partial D_e}{\partial x_{e'}}=T_1\mathbb{I}_{\{e'\in \mathcal{N}(e)\}}\prod_{e'' \in\mathcal{N}(e)\setminus e'}(1-T_1 x_{e''})\,,
\end{split}
\end{equation}
where $\mathbb{I}$ is the indicator function giving one for true arguments and zero for false. At the zero fixed point ($x_e=0$ for all $e$) we find that the Jacobian matrix is the same for both systems,
\begin{equation}
\frac{\partial U_e}{\partial x_{e'}}\bigg|_{\bm{x}=\bm{0}}=\frac{\partial D_e}{\partial x_{e'}}\bigg|_{\bm{x}=\bm{0}}=T_1B_{e,e'}\,,
\end{equation}
where $B$ is the non-backtracking matrix. Major outbreaks are therefore only possible if $T_1> T_c=1/|\lambda_{\max}(B)|$, where $T_c$ is the critical value for the percolation transition. Note that the transition point is the same for all nodes, regardless of any heterogeneity in the network. As shown in \cite{Karrer2014,Hamilton2014}, the role of $\lambda_{\max}(B)$ is a general result for percolation processes on sparse networks. 

\section{The role lifetime distribution}
\label{memory}
As briefly mentioned earlier, although the probability of a major outbreak occurring depends on the detail of the lifetime distribution of the disease, the chance of a particular node catching the disease does not. Two different diseases may have the same transmissibility $T_1$, but one may be more dangerous than the other by virtue of the fact that it is more likely to cause a major outbreak. For fixed transmissibility, which lifetime distribution $\gamma(t)$ is the most dangerous? 
 
To maximise the chance of an outbreak occurring, the disease must guard against the possibility of it dying out in the early stages of its spread. This is achieved by having a \emph{deterministic} infectious period 
\begin{equation}
\gamma(t)=\delta\left(t-\frac{1}{\beta}\log\left[\frac{1}{1-T_1}\right]\right)\,.
\end{equation}
Mathematically, this follows from Jensen's inequality. Moving the power of $n$ outside the integral in equation (\ref{defT}) we find that for $n\geq2$ we have $T_n\geq (T_1)^n$. From the cavity equations (\ref{cav_up}) we then deduce
\begin{equation}
\begin{split}
r_j^{(i)}&\leq1-\prod_{l\in\mathcal{N}(j)\setminus i}\left(1-T_1 r_l^{(j)}  \right)\,,
\end{split}
\label{r_leq}
\end{equation}
with equality if and only if $\gamma(t)$ is a delta function. This result was mentioned in \cite{Trapman2007}, and goes back to older work using percolation theory \cite{Cox1988}.
\begin{figure}
\includegraphics[width=250pt]{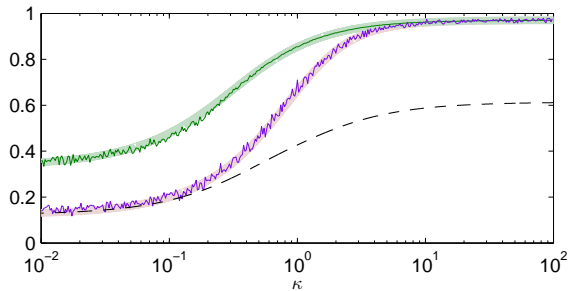}
\caption{Outbreak probability and final size for epidemics with $\beta=1$ and Weibull-distributed lifetimes, on a random 4-regular network with $N=500$ nodes. Thin dark lines show the results of simulations for the fraction of outbreaks affecting more than 10\% of the populations always (lower, purple) and the average final size of those outbreaks (upper, green). Fat pale lines correspond to the risk and vulnerability predictions of the cavity method. The dashed line shows the theoretical lower bound for risk, computed using equation (\ref{r_geq}). The centre line $\kappa=1$ corresponds to Markov dynamics.}
\label{Wei}
\end{figure}

We may also ask how low the probability of an outbreak can get when $T_1$ is fixed. It follows immediately from the definition (\ref{defT}) that the sequence $\{T_n\}$ is positive and decreasing. Thus
\begin{equation}
\begin{split}
r_j^{(i)}&\geq T_1\bigg(1-\prod_{l\in \mathcal{N}(j)\setminus i} \Big(1-r_l^{(j)}\Big)\bigg)\,.
\end{split}
\label{r_geq}
\end{equation}
This rough bound gives the intuition that the risk is minimal when the decay rate of $\{T_n\}$ is also minimal. To achieve this requires a lifetime distribution that is sharply peaked at zero and has a heavy tail. 

The Weibull family of distributions \cite{Weibull1951} describe the time to failure in systems depending on several components, and are a natural choice to model non-Markov disease lifetime distributions (as used in \cite{Cauchemez2011}, for example). In fact, they nicely illustrate the full range of behaviours between the upper and lower bounds given above. Figure \ref{Wei} shows the results of simulations of epidemics with  Weibull-distributed lifetimes, $\gamma(t)=\kappa e^{-t^\kappa}t^{(\kappa-1)}$, where the parameter $\kappa$ controls the shape of the distribution. The infection rate was held constant at $\beta=1$ and the same random 4-regular random network was used for each sample. Small values of $\kappa$ correspond to the hazard rate of the disease getting smaller as it survives longer. This makes the decay of $\{T_n\}$ slower, and the outbreak probability approaches its theoretical minimum. The special case $\kappa=1$ corresponds to Markov disease dynamics (i.e. exponential lifetime distribution). For large $\kappa$ the Weibull distribution approaches a delta function and risk is maximised. 

Finally, we point out that the line for risk in Figure (\ref{Wei}) stays below the line for vulnerability. This is a general fact. Notice that the right-hand-side of equation (\ref{r_leq}) exactly corresponds to the form of the downstream cavity equations (\ref{cav_down}). We can conclude that $v_i$ provides an upper bound for $r_i$. That is to say, the probability that node $i$ will cause a major outbreak if they are the source is always less than their chance of being infected by an outbreak that starts elsewhere. 

\section{Vulnerability and risk ranking}
\label{ranking}
\begin{figure}
\includegraphics[width=250pt]{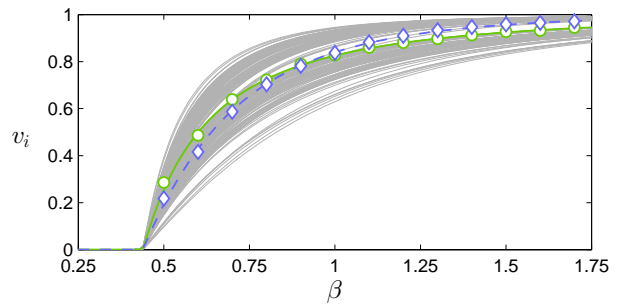}
\caption{node vulnerability as a function of $\beta$ in a random network of $N=10^3$ nodes with degrees five and three (in $50/50$ ratio), generated by the configuration model, with Markov ($\gamma(t)=e^{-t}$) disease dynamics. Grey lines show the result of the cavity equations, with nodes number 10 and 75 highlighted in solid green and dashed blue, respectively. Circles and diamonds show the results of stochastic simulations for these two nodes, averaged over $10^4$ samples. }
\label{vul_rank}
\end{figure}
We know from the previous section that the vulnerability of a node always exceeds its risk, but how do the risks and vulnerabilities of different nodes in the same network compare? For a given network and choice of $\beta$ and $\gamma(t)$, the cavity equations can be used to determine $r_i$ and $v_i$ for every node, and this information may be used to rank the nodes according to the risk they pose to the network, or the risk the network poses to them. As we will see, these rankings depend in detail on the nature of the disease. 

Let us begin by examining vulnerability ranking. node vulnerability depends on the disease specification only through $T_1$, so for simplicity we consider Markov dynamics, fixing $\gamma(t)=e^{-t}$. To generate Figure \ref{vul_rank}, a single random network of $N=10^3$ nodes was created using the configuration model, and the vulnerability of its nodes computed for various $\beta$ using the cavity method. Two things are immediately noticeable from the plot: (i) there is a great deal of heterogeneity between nodes, and (ii) the rank order of vulnerability is not preserved as $\beta$ is varied. In particular, node 10 is more vulnerable than node 75 to less virulent diseases, yet is less vulnerable to more virulent ones. 

To help explain this counter-intuitive finding, we explore the behaviour of the cavity equations near the limits of strongly infectious ($T_1=1$) and weakly infectious ($T_1=T_c$) diseases. Differentiating the vulnerability equation (\ref{vul}) with respect to $T_1$, we have
\begin{equation}
\partial v_i=\sum_{j\in \mathcal{N}(i)}\big(v_j^{(i)}+T_1\partial v_j^{(i)} \big)\prod_{l\in\mathcal{N}(i)\setminus j}\big(1-T_1v_l^{(i)}\big)\,,
\label{vul_diff}
\end{equation}
where $\partial$ is used as short-hand for $\partial/\partial T_1$. Now, as $T_1\to1$ we have $v_j^{(i)}\to1$ also, therefore the product above gives zero unless it is empty. That is, $\partial v_i\to0$ as $T_1\to1$ unless $i$ has only one neighbour. In general, the first non-zero derivative of $v_i$ at $T_1=1$ is 
\begin{equation}
\partial^{|\mathcal{N}(i)|} v_i=|\mathcal{N}(i)|!\prod_{j\in\mathcal{N}(i)}\big(1+\partial v_j^{(i)}\big)\,.
\end{equation}
Similarly, we differentiate the downstream cavity equations (\ref{cav_down}) to find that, 
at $T_1=1$, $\partial v_j^{(i)}$ is only non-zero if $j$ has exactly one other neighbour $l\neq i$, in which case $\partial v_j^{(i)}=1+\partial v_l^{(j)}$. Tracing a path away from $i$ in the direction of $j$ we must eventually, after $L_{ij}$ steps reach a node with degree not equal to two (recall that we ignore the possibility of cycles), whose cavity derivative in either case will be zero. We conclude that $\partial^{|\mathcal{N}(i)|} v_i=|\mathcal{N}(i)|!\prod_{j\in\mathcal{N}(i)}L_{ij}$.

Turning attention now to the percolation transition, we set $T_1= T_c$, where we have $v_i=v_i^{(j)}=0$. Therefore, from (\ref{vul_diff}),
\begin{equation}
\partial v_i=\sum_{j\in \mathcal{N}(i)}T_1\partial v_j^{(i)}=\sum_{j\in \mathcal{N}(i)}T_1\sum_{l\in \mathcal{N}(j)\setminus i}T_1\partial v_l^{(j)}=\ldots\,.
\end{equation}
To make progress, let us suppose that there is some bulk network $B$, which we assume has no special structure and we have solved the cavity equations for that network. To find the value of $\partial v_i$, we must carry out the expansion above for all paths $p$ from $i$ to the bulk $B$. We may write
\begin{equation}
\partial v_i\approx\sum_{p:i\to B}T_c^{|p|}\partial v_B\,,
\end{equation}
where $v_B$ is the mean cavity vulnerability in the bulk. 

Bringing the above results together, we have for each node $i$ a pair of expansions,
\begin{equation}
\begin{split}
&v_i\approx1-(1-T_1)^{|\mathcal{N}(i)|}\prod_{j\in\mathcal{N}(i)}L_{ij}\Bigg.\,,\\
&v_i\approx(T_1-T_c)\sum_{p:i\to B}T_c^{|p|}\partial v_B\,.
\end{split}
\end{equation}
The qualitative insight provided by these equations is the following: vulnerability to highly infectious diseases is mainly controlled by the number of neighbours a node has, whereas vulnerability to weakly infectious diseases depends more subtly on the number and length of paths connecting the node to the infected bulk. The puzzling case of nodes 10 and 75 in Figure \ref{vul_rank} is explained thus: node 75 has higher degree (five, compared to three), yet is the source of just 91 paths of length three, compared to 136 for node 10. We can therefore expect node 10 to more vulnerable close to $T_c$, but less vulnerable for large $T_1$.

\begin{figure}
\includegraphics[width=250pt]{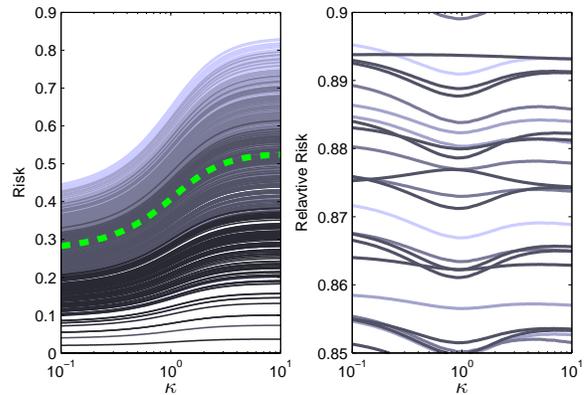}
\caption{Left: node risk as a function of $\kappa$, with $\beta$ varied so as to hold constant $T_1=1/2$. Each solid line corresponds to a different node, shaded according to degree (lighter shades are higher degree); the dashed line gives the mean risk $\bar{r}$. Right: close up of relative risk $r_i/\bar{r}$ for several nodes; crossing lines correspond to changes in the risk ranking.}
\label{Rel_Wei}
\end{figure}

As one might expect in light of the above, the ranking of nodes according to their risk $r_i$ is also not preserved under varying $\beta$. More interestingly, risk rankings are additionally sensitive to the memory characteristics of the disease. Figure \ref{Rel_Wei} shows the result of the cavity equations in a random network with maximum degree five, for Weibull distributed infectious periods with $\kappa$ varying across two orders of magnitude. This time $\beta$ has also been varied simultaneously in order to keep $T_1$ fixed. Although node risk is broadly correlated with degree, this is not a hard rule: there is considerable variation, and many changes to the risk ranking occur as $\kappa$ is varied (even though node vulnerability does not change since $T_1$ is held constant). Curiously, Markov dynamics $\kappa=1$ appear to represent an extreme case, with most nodes experiencing either their highest or lowest relative risk at that point.

\section{Conclusion} 

In the work presented above we have seen how the cavity method can be used to calculate measures of vulnerability and risk in network epidemic models. The vulnerability of a node to an ongoing outbreak is recovered from the solution of the downstream cavity equations (\ref{cav_down}), while the risk of an outbreak occurring with a given starting node is found by the upstream cavity equations (\ref{cav_up}). Node vulnerability was found to be independent of the details of the disease lifetime distribution, however, some nodes are more vulnerable than others to weaker infections, yet less vulnerable to stronger ones. This fact can be understood by an analysis of cavity equations with parameters in the neighbourhood of percolation and complete infection. 

The story for node risk is even more complex, as it depends on the memory characteristics of the disease in a non-trivial way. In particular, we saw how risk is maximised by diseases with deterministic lifetime distributions. This result suggests that real-world diseases, which are subject to evolutionary pressure to optimise their chance of spreading, should have infectious periods that are much less variable than the equivalent exponential distribution, and indeed this appears to be the case \cite{Lloyd2001}. This dependence on disease memory also carries over to the node risk rankings, as seen in Figure \ref{Rel_Wei}.

The main drawback of the method is, of course, the reliance on a tree approximation. Although the method has performed well for all the networks considered in this article (despite most of them containing many cycles), accurate predictions cannot be made for networks with specific structure that favours the existence of cliques or short cycles. This is a common problem in the study of epidemics on networks, which is usually tackled via `moment closure' methods for differential equations \cite{Keeling1999,Rogers2011}. Indeed an explicit link has been made between the cavity method and a particular moment closure scheme \cite{Wilkinson2014}.

Looking to the future, three interesting questions remain unanswered: (i) why do node risk rankings depend on the memory properties of the disease? (ii) why do Markov dynamics extremise relative risk? (iii) how might the results described here be modified in networks with high local clustering? Beyond these problems, several straightforward extensions to the work described here are possible. One may choose to study non-Markov infection rates, although this only has the effect of changing the definition of $T_n$. More interestingly, fully time-dependent cavity equations can be derived to give a more detailed view of the disease progression (as was originally done for vulnerability in \cite{Karrer2010}), and recent work has shown how the method may be used to infer the origin of an outbreak \cite{Lokhov2014}. For added realism in applications, further heterogeneity can be introduced by allowing $\beta$ and $\gamma$ to vary across the network. More generally, the cavity method is widely applicable to other models of cascades on networks, for example, meme spread in social networks or systemic risk in financial markets.  

\textit{Acknowledgements}\par
The author acknowledges funding from the Royal Society, and thanks Dick James, Nick Britton, Thomas House and Reimer K\"{u}hn for useful discussions.


\end{document}